\documentstyle[12pt,A4]{article}
\catcode`\@=11

\@addtoreset{equation}{section}
\def\theequation{\arabic{section}.\arabic{equation}}

\def\@normalsize{\@setsize\normalsize{15pt}\xiipt\@xiipt
\abovedisplayskip 14pt plus3pt minus3pt%
\belowdisplayskip \abovedisplayskip
\abovedisplayshortskip  \z@ plus3pt%
\belowdisplayshortskip  7pt plus3.5pt minus0pt}

\def\small{\@setsize\small{13.6pt}\xipt\@xipt
\abovedisplayskip 13pt plus3pt minus3pt%
\belowdisplayskip \abovedisplayskip
\abovedisplayshortskip  \z@ plus3pt%
\belowdisplayshortskip  7pt plus3.5pt minus0pt
\def\@listi{\parsep 4.5pt plus 2pt minus 1pt
            \itemsep \parsep
            \topsep 9pt plus 3pt minus 3pt}}

\def\underline#1{\relax\ifmmode\@@underline#1\else
        $\@@underline{\hbox{#1}}$\relax\fi}
\@twosidetrue

\relax

\catcode`@=12

\catcode`\@=11

\def\section{\@startsection{section}{1}{\z@}{3.5ex plus 1ex minus
   .2ex}{2.3ex plus .2ex}{\large\bf}}

\def\thesection{\Roman{section}.}

\def\appendix{\setcounter{section}{0}
        \def\thesection{APPENDIX }
        \def\theequation{\Alph{section}.\arabic{equation}}}

\def\FERMIPUB{}

\def\ps@headings{\def\@oddfoot{}\def\@evenfoot{}
\def\@oddhead{\hbox{}\hfill
        \makebox[.5\textwidth]{\raggedright\ignorespaces --\thepage{}--
        \hfill {\rm FERMILAB--Pub--\FERMIPUB}}}
\def\@evenhead{\@oddhead}
\def\subsectionmark##1{\markboth{##1}{}}
}

\catcode`\@=12

\relax

\def\figcap{\section*{Figure Captions\markboth
        {FIGURECAPTIONS}{FIGURECAPTIONS}}\list
        {Fig. \arabic{enumi}:\hfill}{\settowidth\labelwidth{Fig. 999:}
        \leftmargin\labelwidth
        \advance\leftmargin\labelsep\usecounter{enumi}}}
 \relax
\def\tablecap{\section*{Table Captions\markboth
        {TABLECAPTIONS}{TABLECAPTIONS}}\list
        {Table \arabic{enumi}:\hfill}{\settowidth\labelwidth{Table 999:}
        \leftmargin\labelwidth
        \advance\leftmargin\labelsep\usecounter{enumi}}}
 \relax
\def\reflist{\section*{References\markboth
        {REFLIST}{REFLIST}}\list
        {[\arabic{enumi}]\hfill}{\settowidth\labelwidth{[999]}
        \leftmargin\labelwidth
        \advance\leftmargin\labelsep\usecounter{enumi}}}
 \relax

\catcode`\@=11

\def\FERMIPUB{}

\def\ps@headings{\def\@oddfoot{}\def\@evenfoot{}
\def\@oddhead{\hbox{}\hfill
        \makebox[.5\textwidth]{\raggedright\ignorespaces --\thepage{}--
        \hfill {\rm FERMILAB--Pub--\FERMIPUB}}}
\def\@evenhead{\@oddhead}
\def\subsectionmark##1{\markboth{##1}{}}
}

\ps@headings

\relax
\newskip\humongous \humongous=0pt plus 1000pt minus 1000pt
\def\caja{\mathsurround=0pt}
\def\eqalign#1{\,\vcenter{\openup1\jot \caja
        \ialign{\strut \hfil$\displaystyle{##}$&$
        \displaystyle{{}##}$\hfil\crcr#1\crcr}}\,}
\newif\ifdtup

\def\beq{\begin{equation}}
\def\eeq{\end{equation}}

\def\beqn{\begin{eqnarray}}
\def\eeqn{\end{eqnarray}}
\relax

\def\G2{{\; \rm GeV/}c^2}
\def\G{\; \rm GeV}

\def\dotx{\dotx{\dot\overline{x}}}

\relax

\begin{document}
\hbadness=10000
\begin{titlepage}
\nopagebreak
\begin{flushright}

        {\normalsize
 Kanazawa-93-08\\

  September,1993   }\\
\end{flushright}
\vfill
\begin{center}
{\large \bf Study of Minimal String Unification \break
in $Z_8$ Orbifold Models }
\vfill
{\bf Hiroshi Kawabe, Tatsuo Kobayashi } and
{\bf Noriyasu Ohtsubo$^*$}\\

\vspace{1.5cm}
       Department of Physics, Kanazawa University, \\
       Kanazawa, 920-11, Japan \\
 and \\
$^*$Kanazawa Institute of Technology, \\
Ishikawa 921, Japan
\vfill

\end{center}

\vfill
\nopagebreak
\begin{abstract}
We study the construction of the minimal supersymmetric
standard model from the $Z_8$ orbifold models.
We use a target-space duality anomaly cancellation and a
 unification of gauge couplings as constraints.
It is shown that some models obtained through a systematical search realize
the unification of SU(3) and SU(2) coupling constants.

\end{abstract}

\vfill
\end{titlepage}
\pagestyle{plain}
\newpage
\voffset = -2.5 cm

Superstring theories are promising candidates for unified theories of all the
known interactions including gravity.
We could write several types of scenarios from the string scale
$M_{\rm string}$ to the low energy scale.
Although some of them have intermediate scales such as SU(5) and SO(10) GUT's,
 a scenario without the intermediate scales is simplest.
We are much more interested in a minimal string vacuum connected at the string
scale directly to the minimal supersymmetric standard model (MSSM),
 which has the SU(3)$\times$SU(2)$\times$U(1) gauge group, three generations,
a pair of Higgs particles and their superpartners.
Actually, within the framework of the $Z_N$ orbifold models \cite{ZNOrbi}
the MSSM massless spectra with extra matter fields have been obtained in $Z_3$
 and $Z_7$ models [2-7].

Recent LEP measurements [8-11] indicate that gauge couplings of SU(3),
SU(2) and U(1) are simultaneously unified at $M_{\rm GUT}=10^{16}$GeV within
the framework of the MSSM, while in the string theory all the tree level
couplings are identical at
$M_{\rm string}=5.27\times g_{\rm string}\times 10^{17}$GeV
\cite{Kaplunovsky,Derendinger}, where $g_{\rm string}\simeq 1/\sqrt 2$ is the
universal string coupling constant.
The discrepancy between $M_{\rm GUT}$ and $M_{\rm string}$ seems to reject the
 minimal string vacuum.
This situation, however, could be changed when we take into account threshold
corrections due to towers of massive modes.

Recently, the threshold corrections in the orbifold models have been calculated
 explicitly in ref.~[12-15], where the target-space duality plays an important
role.
The duality symmetry is the proper `stringy' feature \cite{Kikkawa,Sakai}.
Here we study duality invariant vacua.
In general the six-dimensional orbifolds have SL(2,${\bf Z})^3$ as the duality
symmetry \cite{Djikgraaf,Shapere}.
Each complex plane of the orbifold has SL(2,${\bf Z})$ duality symmetry.
Loop effects could make this duality symmetry anomalous.
In order to obtain consistent field theories, this duality anomaly should be
cancelled by the Green-Schwarz (GS) mechanism \cite{GS} and loop contributions
from the towers of massive modes.

In ref.~\cite{Ibanez} the constraint of the duality anomaly cancellation was
considered
systematically and the threshold corrections were estimated explicitly.
These analyses have a power to discard a great deal of hopeless models.
Actually, it was shown without exhausting the models that all $Z_N$
orbifold models except $Z_6$-II and $Z_8$-I have no candidate of the MSSM
possessing the consistent couplings with the measurements and $Z_8$-I orbifold
models have a wider range for the promising models than $Z_6$-II ones.
In addition, the $Z_8$-I orbifold has a simpler structure than $Z_6$-II,
because the former has only two independent Wilson lines (WL's) of orders
two, while the latter has three independent WL's \cite{KO1,KO2}.
Therefore we study the possibility for a minimal string unification through the
 threshold effects by examining the $Z_8$-I orbifold models explicitly in this
paper.

The content of the paper is as follows.
First of all we review the construction of the $Z_8$-I orbifold models and then
 investigate systematically massless spectra of the orbifold models in order to
 find the same matter content as the MSSM.
In general, models obtained from the string vacua have an anomalous U(1)
symmetry.
Since such vacua are not stable, the U(1) symmetry breaks through the Higgs
mechnism and some matter fields obtain large masses preserving the $N=1$
supersymmetry \cite{Z3Casas2,Font}.
After this breaking, the massless spectrum is expected to coincide with the
MSSM matter content.
Secondly we consider this possibility and we search the models with the MSSM
matter content plus some pairs of SU(3) triplets (3,1) and $({\overline 3},1)$
as the string massless spectrum.
Next, among the obtained models we investigate the possibility of selecting
out vacua without the duality anomaly through the anomalous U(1) breaking.
Then we study whether the allowed models are able to unify the gauge couplings
of SU(3) and SU(2) so as to be consistent with the mesurements through
renormalization group (RG) equations.

At the begining, we briefly survey the construction of $Z_8$-I orbifold
model, whose 6-dim compact space is obtained by dividing ${\bf R}^6$ in terms
of space group elements $(\theta ,e_i)$.
Here the vector $e_i$ is placed on an SO(9)$\times $SO(5) lattice and the twist
 $\theta$ is its automorphism whose eigenvalues are $(1,2,-3)/8$.
The orbifold models have right-moving RNS and left-moving gauge parts.
Their momenta, $p^t$ and $P^I$, lie on SO(10) and E$_8\times $E$'_8$ lattices,
respectively.
The twist is embedded into the SO(10) lattice as a shift $v^t=(1,2,-3,0,0)/8$.
A shift  vector $V^I$ and WL's $a^I_i$ ($I=1\sim 16$) on the E$_8\times
$E$'_8$  lattice $\Lambda_{{\rm E}_8\times {\rm E'_8}}$ are accompanied with
$\theta $ and $e_i$, respectively.
Refs.~\cite{KO1,KO2} show that each lattice of SO(9) and SO(5) allows only one
independent WL of order two, $i.e.$,
$$2a_1=2a_6=0,\quad a_1=a_2=a_3=a_4,\quad a_5=0 \quad {\rm mod ~}
\Lambda_{{\rm E}_8\times {\rm E}'_8}.
\eqno(1)$$
The shifts also have to fulfill $8V^I=0 \ {\rm mod ~}
\Lambda_{{\rm E}_8\times {\rm E}'_8}$.
The independent shifts of the $Z_8$ orbifold are exhibited at the table VII of
ref.
\cite{Gauge}.
The modular invariance requires the following conditions:
$$ 8\sum_I (a^I_i)^2={\rm integer},\quad 8\sum_I V^I a^I_i={\rm integer},
$$
$$ 8\left( \sum_I (V^I)^2-{7 \over 32} \right)={\rm even} .
\eqno(2)$$

Physical states are classified into untwisted states and twisted ones.
The untwisted states are closed on the torus and their E$_8\times $E$'_8$
momenta $P^I$ fulfill $P^Ia^I_i={\rm integer}$.
They contain gauge bosons and untwisted matters.
The gauge bosons satisfy $P^IV^I=$integer and the matters belong to
$P^IV^I=1/8,2/8,5/8$ (mod integer).
The twisted states are closed on the orbifold and they are invariant under the
$\theta ^k$ twist, where $k$=1,2,4,5.
They are associated with fixed points, which are represented by
the space group elements ($\theta ^k,n_ie_i$).
The $k\geq 2$ twisted states attached to each fixed point are not always
invariant under the $\theta$ twist.
In order to obtain $\theta$-eigenstates, we must take linear combinations of
those states,
\footnote{See in detail ref.~\cite{KO3,KO1,KO2}.}
whose eigenvalues under $\theta$ are denoted by $\gamma$ hereafter.
The twisted states on the fixed points $(\theta^k,n_ie_i)$ have
E$_8\times $E$'_8$ momenta $P^I+kV^I+n_ia^I_i$.
Massless states with the momenta have to satisfy the following condition,
$$ {1\over 2}\sum_I (P^I+kV^I+n_i a_i^I)^2+N_k -1+c_k =0 ,
\eqno(3)$$
where $N_k$ is a number operator and $c_k$ is obtained as
$$ c_k={1\over 2}\sum_{t=1}^3 \left(|kv^t|-{\rm Int}(|kv^t|)\right)
\left(1-|kv^t|+{\rm Int}(|kv^t|)\right),
\eqno(4)$$
where ${\rm Int}(a)$ represents an integer part of $a$.
The states with the momenta \break ($p^t+kv^t,P^I+kV^I+n_ia^I_i$) have the
following GSO phases:
$$\eqalign{
 \Delta =& P^{(k)}\gamma{\rm exp}[2\pi i\left( -{1\over 2k}\sum_I
(kV^I+n_i a_i^I)^2+{k\over 2}\sum_t (v^t)^2 \right.\cr
 &\left.+{1\over k}\sum_I(kV^I+n_i a_i^I)(P^I+kV^I+n_i a_i^I)
-\sum_tv^t(p^t+kv^t) \right) ],\cr}
\eqno(5)$$
where $P^{(k)}$ is the $Z_8$ phase factor from oscillator contributions.
The physical states should satisfy $\Delta $=1.
Degeneracy numbers of the massless states in the twisted sectors are exhibited
in refs.\cite{KO1,KO2}.

Our aim is to obtain the models which have just the same matter content as the
MSSM.
We search models with the gauge group
SU(3)$\times$SU(2)$\times$U(1)$^5$ in the observable sector.
Following refs.~\cite{Reduce,Z7Casas}, we fix eight SU(3)$\times$SU(2)
non-zero roots as $P^I=(0,0,$\underline{1,$-1$,0}$,0,0,0)$ and
(\underline{1,$-1$}$,0,0,0,0,0,0)$,
where the underlines represent arbitrary permutations.
We investigate the massless spectra (combinations of the shift and the WL's)
 of
$$\eqalign{  \cdot & {\rm \ the \
untwisted \ matters \ in \ the \ observable \ sector }\cr
&{\rm \ (observed \ elements \ of \ the \ shifts \ }V^I\mbox {\rm \ and \ the
\  WL's \ } a^I\  (I=1\sim 8)){\rm \  at \ Stage \ 1},
\cr
\cdot & \mbox {\rm \ the \ twisted \ matter \ with \ vanishing \ WL's  }\cr
&{\rm \ (whole \ elements \ of \ the \ shifts \ }V^I \ (I=1\sim 16)) {\rm \ at
\ Stage \ 2, }\cr
\cdot & {\rm \ the \ other \ twisted \ matters \ and \ the \ untwisted
\ matters \ in \ the \ hidden \ sector }\cr
&{\rm \ (the \ whole \ elements \ of \ the \ shifts \ }V^I \mbox {\rm \ and \
the \ WL's \ } a^I\ (I=1\sim 16)) {\rm \ at \ Stage \ 3.}\cr}
$$
At Stage 1, we select combinations of the shift and the WL's which induce just
eight non-zero roots as expressed above and no extra matter of $(\overline{3},
2)$ or (3,1) representation of SU(3)$\times$SU(2).
Under this selection rule, twenty shifts remain and they belong to
No.15,16,20,22,23,24,25,26 and 29 of the table VII of ref.~\cite{Gauge}.
Each shift has 1$\sim$4 types of WL's as allowed combinations.
At Stage 2 we impose a nonexistence condition of ($\overline{3},2$) and (3,1)
matters upon the massless spectra in order to choose the hidden elements of the
 shifts $V^I$ $(I=9\sim 16)$ corresponding to the observable ones selected in
Stage 1.
We obtain allowed combinations as follows,
$$\mbox {\rm (Shift, WL's)=(I,i),(I,ii),(II,i),(II,ii),(III,iii),(IV,iii),
(IV,iv)},
\eqno(6)$$
where the shifts are given by
$${\rm I}:V=(1,1,2,2,2,3,2,-1;4,4,3,3,1,1,0,0)/8, \hspace{2in}$$
$${\rm II}:V=(1,1,2,2,2,3,2,-1;3,3,2,2,2,2,1,-1)/8, \hspace{2in}$$
$${\rm III}:V=(2,2,2,2,2,2,1,-1;3,3,3,3,2,2,2,-2)/8, \hspace{2in}$$
$${\rm IV}:V=(2,2,2,2,2,3,2,-1;3,3,3,2,2,2,2,-1)/8, \hspace{2in}
\eqno(7)$$
and the observable elements of the WL's are given by
$${\rm i}:a_1=(0,0,0,0,0,0,2,2)/4 ~~ a_2=(1,1,-1,-1,-1,1,-1,1)/4,$$
$${\rm ii}:a_1=(0,0,2,-2,-2,-2,0,0)/4 ~~ a_2=(1,1,3,-1,-1,1,1,-1)/4,
$$
$${\rm iii}:a_1=(0,0,2,-2,-2,0,-2,0)/4 ~~ a_2=(1,1,3,-1,-1,-1,1,1)/4
,$$
$${\rm iv}:a_1=(0,0,2,-2,-2,0,0,2)/4 ~~ a_2=(1,1,3,-1,-1,-1,-1,-1)/4
.\eqno(8)$$
At Stage 3 we have searched the models which have three or more
generations and no ($\overline{3}$,2), but we have not been able to obtain such
 models for all combinations of eq.~(6).

By the above result, however, we can not conclude that the minimal string model
 is not obtained from the $Z_8$-I orbifold models.
The matters ($\overline{3}$,2) or (3,1) are possible to be included in the
massless spectra, because they might couple with extra (3,2) or
$(\overline{3},1)$ matters and obtain heavy masses through
the anomalous U(1) breaking \cite{Z3Casas2,Font}.
Now, we permit the existence of some pairs of (3,1) and $(\overline{3},1)$
matters in the twisted sector by way of trial.
Let us go back to Stage 2.
Then we find each observed shift subjects 4$\sim$9 hidden
shifts.
We have obtained 249 combinations of the shifts $V^I$
($I=1\sim 16$) and the observed elements of the WL's $a^I_i$ $(I=1 \sim 8$).

It is notable that models with larger hidden gauge groups are easy
to analyze and models involving smaller ones have the possibility of mixture
of observed and hidden matters.
Therefore, we pay attention to models with rather large gauge groups.
We have carried out Stage 3 under the condition of non-existence
of $(\overline{3},2)$.
The result is as follows.
The largest hidden gauge group is SO(10)$'\times $U(1)$'^3$, which is realized
in two models named Model~1 and Model~2.
(There are also models with hidden gauge groups SU(6)$'$, SU(5)$'$ and so on.)
Massless contents of the models are obtained as follows,
$$\eqalign{& 3[(3,2)+2(\overline{3},1)+(1,2)+(1,1)]+2(1,2) \cr
& +17[(\overline{3},1)+(3,1)]+34(1,2)+2(10)'+159(1,1) ~~{\rm for \
Model~1,}\cr
& 3[(3,2)+2(\overline{3},1)+(1,2)+(1,1)]+2(1,2) \cr
& +16[(\overline{3},1)+(3,1)]+32(1,2)+2(10)'+159(1,1) ~~{\rm for \
Model~2}.\cr}
\eqno(9)$$
Model 1 is derived from the following shift and the WL's:
$$\quad V=(1,1,2,2,2,3,3,0;2,2,1,1,1,1,1,1)/8$$
$$\quad a_1=(0,0,0,0,0,2,0,2;2,0,-2,0,0,0,0,0)/4,
\eqno(10) $$
$$\quad a_2=(1,1,-3,1,1,-1,1,-1;1,-1,-1,1,1,1,1,1)/4.$$
Model~2 is obtained by the same shift $V^I$ and the same WL $a_1$ as eq.~(10)
and \break
$a_2=(1,1,-3,1,1,-1,1,-1;-3,-1,1,-1,-1,-1,-1,-1)/4$.
Both the models have the anomalous U(1) symmetry, so the vacua are unstable.
We could analyze U(1) charges following to refs.~\cite{Z3Casas2,Font}, and
discuss the breakings of extra U(1) gauge symmetries to obtain stable vacua.
Instead of doing so, we investigate whether the duality invariant vacua with
the MSSM matter content can be obtained from the above two models after some
type of the U(1) breaking occurs and the extra matters become massive
following ref.~\cite{Ibanez}.
One of the reasons for this option is that duality anomaly cancellation
condition is powerful enough to discard lots of hopeless models.

Effective field theories derived from the 4-dim orbifold models must be
invariant under the following transformation of a moduli $T_i$:
$$ T_i \rightarrow {a_iT_i-ib_i \over ic_iT_i+d_i} ,
\eqno(11)$$
with $a_i,b_i,c_i,d_i\in {\bf Z}$ and $a_id_i-b_ic_i=1$.
The moduli $T_i$ is associated with one of three complex planes of the
orbifold.
Under the duality transformation, the matter $A_{\alpha}$
transforms as
$$ A_{\alpha} \rightarrow A_{\alpha} \prod_{i=1}^3(ic_iT_i+d_i)^{n^i_{\alpha}},
\eqno(12)$$
where $n^i_{\alpha}$ is called modular weight which is associated with the
$i$-th plane \cite{Dixon2,Ibanez,Bailin}.
For untwisted matters associated with the $i$-th complex plane, the modular
weights are $n_{\alpha}^j=-\delta^{ij}$.
The $k$=1,2,4,5 twisted matters without oscillator contributions
have the modular weights
$n_{\alpha}$=$(-7,-6,-3)/8$,$(-6,-4,-6)/8$,$(-4,0,-4)/8$,$(-3,-6,-7)/8$,
respectively.
An oscillator $a_i$ of the $i$-th complex plane reduces $i$-th
elements of $n_\alpha$ by one.

The duality anomaly is caused by triangle graphs which have K\"ahler and
curvature connections as one of external lines.
The anomaly can be cancelled by a combination of two ways.
One is GS mechanism, which induces non-trivial duality transformation of the
dilaton field.
The other way is due to the threshold effects for the gauge coupling terms.
The threshold corrections of the gauge coupling constants depend only on the
modulus whose complex planes are not rotated in all the twists, because those
planes have $N=2$ supermultiplets as well as $N=4$ supermultiplets.
Since the first and the third planes of $Z_8$-I orbifold are rotated under all
the twists, the anomalies associated with the planes have to be cancelled only
by the GS mechanism.
On the other hand, the second plane concerning with the $\theta $-eigenvalue
2/8 is fixed under the $\theta^4$ twist, so the threshold correction depends
only on $T_2$.
Both of the GS mechanism and the threshold effects contribute to the
anomaly cancellation about $T_2$.

Actually, anomaly coefficients of the $i$-th plane with respect to SU(3),
 SU(2) and SO(10)$'$ are obtained as
$$ b_{\rm SU(3)}^{\prime i}=-3+\sum_{\alpha \in (3,2)}(2n_{\alpha}^i+1)
+\sum_{\alpha \in (\overline{3},1)}(n_{\alpha}^i+{1\over 2}) ,$$
$$ b_{\rm SU(2)}^{\prime i}=-2+\sum_{\alpha \in (3,2)}3(n_{\alpha}^i+
{1 \over 2})+\sum_{\alpha \in (1,2)}(n_{\alpha}^i+{1\over 2}) ,$$
$$ b_{\rm SO(10)}^{\prime i}=-4+\sum_{\alpha \in (10)'}(n_{\alpha}^i +
{1\over 2}) .
\eqno(13)$$
Since the GS term is gauge invariant, we obtain the anomaly cancellation
condition for completely rotated planes as follows,
$$ b_{\rm SU(3)}^{\prime i}=b_{\rm SU(2)}^{\prime i}
=b_{\rm SO(10)}^{\prime i} \quad (i=1,3),
\eqno(14)$$
where $b_{\rm SO(10)}^{\prime 1}=-39/8$ and
$b_{\rm SO(10)}^{\prime 3}=-27/8$ in both the models.

Model 1 and Model 2 have extra matters other than the MSSM matter content.
Now, we select massless matters remaining in the stable duality invariant
vacua in order to get the MSSM model, that is, we pick up six
($\overline{3}$,1) and five (1,2) as well as three (3,2) from  the matter
content of eq.~(9) under the condition (14).
Table 1$\sim$4 express all the matters of Model 1 except (3,1) and (1,1).
The first columns show the degeneracy numbers of the states.
Types of WL's are found in the fourth columns, where $[n,n']$ denotes
$n=\sum^4_{i=1}n_i$ and $n'=n_6$ (mod 2).
The oscillators $a_i$ involved in the states are shown in the fifth columns.
The last two columns of Table~2 express two types of choice for six
($\overline{3}$,1) matters, $i.e.$ I and II.
The numerator of the columns means the required number to pick up from the
group of matters which belong to the same denominator in a twisted sector, and
the denominator is the number of whole matters which have the same modular
weights for the first and the third planes.
Table 3 expresses three types of choice for five (1,2) matters, $i.e.$
A, B and C, in the same way as Table 2.
Making use of these two tables, we can take all the possible combinations with
respect to the whole modular weights for deriving the duality invariant vacua.
Note that the value of $b^{\prime 2}_{\rm SU(2)}$ depends on whether or not we
choose oscillators in the types B and C.
Alternatively, we consider the case where the two $(10)'$ matters obtain
heavy masses.
In this case we have $b^{\prime i}_{\rm SO(10)}=-4$.
But, we have found no solution of eq.~(14), when we pick up six
($\overline{3}$,1)
and five (1,2) as well as three (3,2) from  the matter content of eq.~(9).

Now, we discuss the one-loop running gauge coupling contants including the
threshold effects as follows,
$$ {1\over g_a^2(\mu)}={1\over g_{\rm string}^2}+{b_a \over 16\pi^2}
{\rm log}{M_{\rm string}^2 \over \mu^2}-{1\over 16\pi^2}
(b_a^{\prime 2}- \delta_{GS}^2){\rm log}[(T_2+\overline{T}_2)|\eta(T_2)|^4],
\eqno(15)$$
where $\eta(T)$ is the Dedekind function, $\delta_{GS}^2$ are gauge group
independent GS coefficients and $b_a$ are $N=1$ $\beta$-function coefficients,
$i.e.$, $b_{\rm SU(3)}=-3$, $b_{\rm SU(2)}=1$ and \break $b_{\rm SO(10)}=-11$.
For the anomaly coefficients of Model 1, the types I and II of SU(3)
parts lead to $b'^2_{\rm SU(3)}=1/4$ and $-7/4$, respectively.
The types A, B and C of SU(2) lead to $b'^2_{\rm SU(2)}=-7/4, -3/4$ and $1/4$
for the choice of no oscillators, respectively, while the type~B including
a $k=1$, $N_2=2/8$ oscillator state leads to $b'^2_{\rm SU(2)}=-7/4$.
Further a choice including a $k=5$, $N_2=2/8$ oscillator state in the
types B and C reduces the above values of $b'^2_{\rm SU(2)}$ by one.
On the other hand, we obtain $b'^2_{\rm SO(10)}=-15/4$ in the case with two
$(10)'$ matters.

Next, we consider the unification of SU(3) and SU(2) gauge couplings.
{}From eq.~(15) the unification mass scale $M_{\rm GUT}$ subjects the following
equation,
$$ {\rm log}{M_{\rm GUT} \over M_{\rm string}}={b_{\rm SU(2)}^{\prime 2}-
b_{\rm SU(3)}^{\prime 2} \over 2(b_{\rm SU(3)}-b_{\rm SU(2)})}{\rm log}
[(T_2+\overline{T}_2)|\eta(T_2)|^4] .
\eqno(16)$$
We further select the combinations of types of SU(3) and SU(2) by the condition
 $M_{\rm GUT}<M_{\rm string}$.
This condition is equivalent to $b'^2_{SU(3)}>b'^2_{SU(2)}$ because the inside
of the square brackets in eq.~(16) is always smaller than 1.
As the results, there are four combinations of the allowed values:
$$(b'^2_{\rm SU(3)},b'^2_{\rm SU(2)})=(1,-7)/4,\ (1,-3)/4,\ (1,-11)/4,\
(-7,-11)/4.
\eqno(17)$$
The similar results are derived in Model 2.
Under the condition $M_{\rm GUT} \sim M_{\rm string}/37$, values of the
differences $b'_{\rm SU(3)}-b'_{\rm SU(2)}=8/3,4,8$ lead
to Re$T_2 \sim12,17,31$, respectively.

Further, we also study the RG flow of the SO(10)$'$ gauge coupling.
The gaugino condensation of the hidden group SO(10)$'$ might lead to a
realistic SUSY-breaking,
\footnote{See e.g. \cite{Nilles} and the references therein.}
although the dynamics of the condensation has never been understood yet.
The scale of the condensation $M_{\rm COND}$ and that of the observable
SUSY-breaking $M_{\rm SUSY}$ are related as
$M_{\rm SUSY} \simeq M_{\rm COND}^3/M_{\rm P}^2$, where $M_{\rm P}$ is the
Planck scale.
So the condensation must happen near by $10^{13}$ GeV in order to derive the
SUSY-breaking at 1 TeV.
For the allowed four combinations of eq.~(17), we impose
$\alpha ^{-1}_{\rm GUT}=25.7$ ($\alpha=g^2/4\pi$) on eq.~(15) in order to
draw the RG flow of the SO(10)$'$ coupling constant.
We find that the combination $(b'^2_{\rm SU(3)},b'^2_{\rm SU(2)})=(1,-3)/4$
 of eq.~(17) leads to $\alpha^{-1}_{\rm SO(10)}=0$ at $10^{13}$GeV.
Therefore, in the model with this combination the condensation might happen
at a higher energy than $10^{13}$GeV.
On the other hand, the combination $(b'^2_{\rm SU(3)},b'^2_{\rm SU(2)})=
(1,-11)/4$ and the other two lead to $\alpha^{-1}_{\rm SO(10)}=5.9$ and 4.4
at 10$^{13}$GeV, respectively.
In these cases, the condensation might happen near by $10^{13}$GeV.

At last we attempt to assign the representations of Model 1 to the
matter superfields of the MSSM.
Let us take the type I of Table 3 and the type C of Table 2 as an example.
Suppose that we assign quarks ($Q_i$,$U_i$,$D_i$, ($i=1,2,3)$), a pair of
Higgs particles ($\tilde H$,$H$) and lepton doublets ($L_i$, ($i=1,2,3)$) as
shown in the tables.
Then these particles have correct hypercharges under the following basis:
$$Y=(2U_2-9U_3-3U_4-3U_5+3{U'}_1-2{U'}_2+16{U'}_3)/48,\eqno(18)$$
where
$$\begin{array}{rclrcl}
U_1&=&(1, 1, 0, 0, 0, 0, 0, 0)/2,& \qquad U_2&=&(0, 0, 1, 1, 1, 0, 0, 0)/2,\\
U_3&=&(0, 0, 0, 0, 0, 1, 0, 0),& \qquad U_4&=&(0, 0, 0, 0, 0, 0, 1, 0),\\
U_5&=&(0, 0, 0, 0, 0, 0, 0, 1),& \qquad {U'}_1&=&(1, 1, 0, 0, 0, 0, 0, 0)',\\
{U'}_2&=&(1, -1, -2, 0, 0, 0, 0, 0)'/2,&\qquad{U'}_3&=
&(1, -1, 1, 3, 3, 3, 3, 3)'/16
{}.
\end{array}\eqno(19)$$
All the U(1) invariances and space group selection rules allow
renormalizable couplings of $Q_iHU_j$ and $Q_i\tilde HD_j$, ($i,j=2,3$).
If a lepton singlet $E_3$ is assigned to a singlet in $k=2$ sector with WL's
[1,1] and U(1) charges $(0,6,-4,-4,6,4,0,0)$, it couples with the lepton $L_3$
as $L_3HE_3$.
The other lepton doublets have no renormalizable coupling with $H$ and all the
singlets.
There are 38 candidates for the first and the second generations of lepton
singlets ($E_1$, $E_2$).
Although the above assignment may be consistent with the low energy
phenomenology, it conflicts with a theoretical requirement.
A sum of the hypercharges of all the massless particles in Model 1
does not vanish
($\sum Y= 21$).

In this letter, we have studied the $Z_8$-I orbifold models with two Wilson
lines systematically in order to construct the MSSM.
It has been shown that the $Z_8$-I orbifold models can not lead to the MSSM
as the string massless spectra.
We have also examined the models which have matter contents of the MSSM with
the extra matters and have obtained two models involving the SO(10)$'$
hidden symmetry.
So as to lead to the MSSM, we have imposed the duality anomaly cancellation
condition on the models.
The unification of SU(3) and SU(2) gauge couplings has also been investigated.
We have found its solutions under some values of $T_2$.
The remaining problem is to show what kinds of symmetry breakings induce such
vacuua.
Further the renomalization group flow of SO(10)$'$ was discussed.
We have tried to assign representation matters of the Model 1 to the
MSSM matters taking into account the hypercharges of them.
We could also study other models which have hidden gauge
groups SU(6)$'$, SU(5)$'$ and so on.
The above approach could be extended to cases with extra matters of the
($\overline{3}$,2) representation.
It is interesting to investigate the $Z_6$-II and $Z_M \times Z_N$ orbifold
models similarly.

\vspace{0.8 cm}
\leftline{\large \bf Acknowledgement}
\vspace{0.8 cm}

The authors would like to thank D.~Suematsu and K.~Matsubara for useful
discussions and S.~Kiura for helpful advices of computor analyses.


\newpage
\pagestyle{empty}
\noindent

\begin{center}

{\large Table 1. (3,2) representations}\\
\vspace{10mm}
\footnotesize
\begin{tabular}{|c|c|c|c|c|r|r|r|r|r|r|r|r||c|}
\hline
Deg. & $k$  & $P^IV^I$ & WL & Osc. & $U_1$ & $U_2$ & $U_3$ & $U_4$ & $U_5$ &
$U'_1$ & $U'_2$ & $U'_3$ & \\ \hline \hline

 1         & 0    &  1/8     & -  &  -         &  $-4$ &    4  &   0   &   0
&   0   &    0   &   0    &   0  & ($Q_1$)  \\ \hline

 2         & 2    &   -      &[1,1]& -         &   0   &    1  &   0   &   0
&   2   &    4   &   0    &   0  & ($Q_2,Q_3$)  \\ \hline
                                   \end{tabular}

\vspace{10mm}
\normalsize
{\large Table 2. $(\bar3,1)$ representations}\\
\scriptsize
\vspace{10mm}
\begin{tabular}{|c|c|c|c|c|r|r|r|r|r|r|r|r||cc|c|}
\hline
Deg.  & $k$  & $P^IV^I$ & WL & Osc. & $U_1$ & $U_2$ & $U_3$ & $U_4$ & $U_5$ &
$U'_1$ & $U'_2$ & $U'_3$ & \multicolumn{2}{c|}{ I}  &  II \\ \hline \hline

   1   &  0   &  1/8     & -  &  -    &  0    & $-4$  &   0   &   8   &   0
&    0   &   0    &   0    &    0 &  &  0      \\

   1   &  0   &  5/8     & -  &  -    &  4    &  2    &   4   &   4   & $-4$
&    0   &   0    &   0    &   1/1 & ($U_1$)  &  1/1    \\

   1   &  0   &  2/8     & -  &  -    &  4    &  2    & $-4$  &   4   &  4    &
  0     &   0    &   0    &    0  &  &  1/1    \\ \hline

   1   &  1   &   -      &[0,0]& -    &  1    &  $-1$ &  $-5$ &   3   &  0    &
   4    &  $-1$  &  1    &  1/3  &  & 1/3     \\

   1   &  1   &   -      &[0,0]& -    &  1    &  $-1$ &  3    &  $-5$ &  0    &
   4    &  $-1$  &  1    &  1/3 & ($D_1$)  & 1/3     \\

   1   &  1   &   -      &[1,1]& -    &  3    &   2   & $-3$  &  $-3$ &  2    &
   0    &   1    &  $-1$ &  1/3 &   & 1/3     \\

   1   &  1   &   -      &[1,1]& $a_1$&  $-1$ &  $-4$ &  1    &   1   & $-2$  &
   0    &   1    &  $-1$ &   0  &   &  0      \\ \hline

   2   &  2   &   -      &[0,1]& -    &  0    &  $-1$ &   0   &  $-4$ &  2    &
   0    & $-6$  &   0    &   0  &   & 2/6     \\

   2   &  2   &   -      &[1,1]& -    &  0    &  $-1$ &  $-4$ &  $-4$ & $-2$  &
   4    &   0    &   0    &   0  &   & 2/6     \\

   2   &  2   &   -      &[1,1]& -    &  0    &  $-1$ &   4   &   4   & $-2$  &
   4    &   0    &   0    &   0  &   & 2/6     \\ \hline

   4   &  4   &   -      &[0,0]& -    &  0    &   2   &   0   &   0   &  4    &
  $-8$  &   0    &   0    &  4/10& ($U_2,U_3$)  & 1/10    \\

   2   &  4   &   -      &[0,0]& -    &  0    &   2   &   0   &   0   &  4    &
   8    &   0    &   0    &  4/10&   & 1/10    \\

   2   &  4   &   -      &[1,0]& -    &  0    &   2   &  $-4$ &   0   &  0    &
 $-4$   &  6    &   0    &  4/10&   & 1/10    \\

   2   &  4   &   -      &[1,0]& -    &  0    &   2   &  $-4$ &   0   &  0    &
   4    & $-6$  &   0    &  4/10&  ($D_2,D_3$) & 1/10    \\ \hline
                                   \end{tabular}

\newpage
\normalsize
{\large Table 3. $(1,2)$ representations}\\
\vspace{10mm}
\scriptsize

\begin{tabular}{|c|c|c|c|c|r|r|r|r|r|r|r|r||c|c|cc|}
\hline
Deg. & $k$  & $P^IV^I$ & WL & Osc. & $U_1$ & $U_2$ & $U_3$ & $U_4$ & $U_5$ &
$U'_1$ & $U'_2$ & $U'_3$ & A & B & \multicolumn{2}{c|}{C} \\ \hline \hline


  1    &  0   &  5/8     &  - &   -   &   0   & $-6$ &  4    &  $-4$ & $-4$  &
   0    &    0   &   0    &   0    &  1/1    &  1/1 & ($L_1$)  \\

  1    &  0   &  2/8     &  - &   -   &  $-4$ &  0    &  0    &   8   &  0    &
   0    &    0   &   0    &   0    &   0    &   0  &    \\

  1    &  0   &  2/8     &  - &   -   &   0   & $-6$ & $-4$  &  $-4$ &  4    &
   0    &  0   &   0      &   0    &   0     &   0  &    \\ \hline

  1    &  1   &   -      &[1,1]&  -   & $-5$ &   0   &   1   &   1   &  $-2$ &
  0     &  1   & $-1$    &  0     &   1/2   &  0   &    \\

  1    &  1   &   -      &[1,1]&$a_2$ &   3   &   0   &   1   &   1   &  $-2$ &
  0     &  1   & $-1$    &  0     &   1/2   &  0   &    \\

  1    &  1   &   -      &[0,0]&$a_1$ &   1   &  $-3$ & $-1$  & $-1$  &   4   &
  4     & $-1$ & 1       &  0     &    0    &  0   &    \\

  1    &  1   &   -      &[1,1]&$(a_1)^2$&  3   &   0   &   1   &   1   & $-2$
  & 0     &   1  & $-1$    &  0     &    0    &  0   &    \\ \hline

  2    &  2   &   -      &[1,0]&  -   &   2   &   0   & $-2$  &  2    &  0    &
  $-4$  & $-4$ & $-2$    & 1/8    &   1/8   & 2/8  &    \\

  2    &  2   &   -      &[1,0]&  -   &   2   &   0   & $-2$  &  2    &  0    &
  $-4$  &  4   &  2      & 1/8    &   1/8   & 2/8  &    \\

  2    &  2   &   -      &[0,1]&  -   &   0   &  $-3$ &  4    &  0    & $-2$  &
   0    & $-6$&   0      & 1/8    &   1/8   & 2/8  & ($H$) \\

  2    &  2   &   -      &[1,1]&  -   &   0   &  $-3$ &  0    &  0    & $-6$  &
   4    &  0   &   0      & 1/8    &   1/8   & 2/8  & ($\tilde H$) \\ \hline

  1    &  4   &   -      &[0,0]&  -   &   0   &   0   &  4    &   4   &  0    &
  $-8$  &  0   &   0      & 1/16   &   0     & 1/16 &    \\

  1    &  4   &   -      &[0,0]&  -   &   0   &   0   &  4    &   4   &  0    &
   8    &  0   &   0      & 1/16   &   0     & 1/16 &    \\

  2    &  4   &   -      &[0,0]&  -   &   0   &   0   & $-4$  & $-4$  &  0    &
  $-8$  &  0   &   0      & 1/16   &   0     & 1/16 &    \\

  4    &  4   &   -      &[0,0]&  -   &   0   &   0   & $-4$  & $-4$  &  0    &
    8   &  0   &   0      & 1/16   &   0     & 1/16 &    \\

  2    &  4   &   -      &[1,0]&  -   &   0   &   0   &  0    & $-4$  &  4    &
  $-4$  & 6   &   0      & 1/16   &   0     & 1/16 &    \\

  2    &  4   &   -      &[1,0]&  -   &   0   &   0   &  0    & $-4$  &  4    &
   4    & $-6$&   0      & 1/16   &   0     & 1/16 &    \\

  2    &  4   &   -      &[1,0]&  -   &   0   &   0   &  0    &  4    & $-4$  &
  $-4$  &  6  &   0      & 1/16   &   0     & 1/16 & ($L_3$) \\

  2    &  4   &   -      &[1,0]&  -   &   0   &   0   &  0    &  4    & $-4$  &
   4    & $-6$&   0      & 1/16   &   0     & 1/16 &    \\ \hline

  1    &  5   &   -      &[1,0]&  -   &   1   &   3   &  3    & $-1$  &   4   &
   0    &  5  &  1      & 3/3    &   2/3   & 1/3  & ($L_2$) \\

  1    &  5   &   -      &[0,1]&  -   & $-1$  &   0   &  1    &   5   &   2   &
   4    & $-5$& $-1$    & 3/3    &   2/3   & 1/3  &    \\

  1    &  5   &   -      &[0,0]& $a_2$&  1    &   3   &  $-1$ & $-1$  &   0   &
 $-4$   & $-1$ &  1      & 3/3    &   2/3   & 1/3  &    \\

  1    &  5   &   -      &[0,0]& $(a_3)^2$& 1   &   3   &  $-1$ & $-1$  &   0
 & $-4$   & $-1$ &  1      &  0     &    0    &  0   &    \\ \hline
                                   \end{tabular}

 \vspace{10mm}
\normalsize
{\large Table 4. Hidden $(10)'$ representations}\\
\vspace{10mm}
\footnotesize

\begin{tabular}{|c|c|c|c|c|r|r|r|r|r|r|r|r|}
\hline
Degen. & $k$  & $P^IV^I$ & WL & Osci. & $U_1$ & $U_2$ & $U_3$ & $U_4$ & $U_5$ &
 $U'_1$ & $U'_2$ & $U'_3$ \\ \hline \hline


  1    &  0   &  1/8     &  - &  -    &   0   &   0   &   0   &   0   &   0   &
    8   & $-4$   &  $-2$ \\ \hline

  1    &  1   &   -      &[0,1]& -    &  $-1$ &   0   &  $-3$ &   1   &   2   &
    4   & $-1$   &   1   \\ \hline
                                   \end{tabular}
\end{center}

\end{document}